\documentclass[aps,floats,showpacs,amssymb,tightenlines]{revtex4}

\usepackage{amsmath}
\usepackage{amsfonts}
\usepackage{amscd}
\usepackage{epsfig}
\usepackage{amssymb}
\usepackage{tabularx}

\def\w{\omega}

\def\ri{r_{i}}
\def\ro{r_{o}}

\begin{document}
\title{Gravitational instabilities in Kerr spacetimes}

\author{Gustavo Dotti, Reinaldo J. Gleiser}
\affiliation {Facultad de Matem\'atica, Astronom\'{\i}a y
F\'{\i}sica (FaMAF), Universidad Nacional de C\'ordoba. Ciudad
Universitaria, (5000) C\'ordoba,\ Argentina}
\author{Ignacio F. Ranea-Sandoval, H\'ector Vucetich}
\affiliation{Facultad de Ciencias Astron\'omicas y Geof\'{\i}sicas, Universidad Nacional de La Plata.
 Paseo del Bosque S/N 1900. La Plata, Argentina}

\begin{abstract}
In this paper we consider the possible existence of unstable axisymmetric modes in Kerr space times, resulting from exponentially growing solutions of the Teukolsky equation. We describe a transformation that casts the radial equation that results upon separation of variables in the Teukolsky equation, in the form of a Schr\"odinger equation, and combine the properties of the solutions of this equations with some recent results on the asymptotic behaviour of spin weighted spheroidal harmonics to prove the existence of an infinite family of unstable modes. Thus we prove that the stationary region beyond a Kerr black hole inner horizon is unstable under gravitational linear perturbations. We also prove that  Kerr space-time with angular momentum larger than its square mass, which has a naked singularity, is unstable.

\end{abstract}

\pacs{04.50.+h,04.20.-q,04.70.-s, 04.30.-w}

\maketitle

\section{Introduction}
In 1916, shortly after Einstein published his vacuum field equations,  Schwarzschild
found a solution that represents what came to be known as a
static {\em black hole}. It took almost fifty years to
find a metric representing a {\em rotating} black hole. This is Kerr's solution
 \cite{kerr}, which,  in Boyer-Lindquist coordinates reads
\begin{equation} \label{kerr}ds^2 = - \frac{(\Delta-a^2 \sin^2 \theta) }{\Sigma} dt^2 -2a\sin^2
\theta \frac{(r^2 + a^2 - \Delta)}{\Sigma}
dt d\phi + \left[ \frac{(r^2+a^2)^2 - \Delta a^2 \sin^2 \theta}{\Sigma} \right] \sin^2 \theta d\phi^2
+
\frac{\Sigma}{\Delta} dr^2 + \Sigma  d\theta^2 ,
\end{equation}
where $\Sigma = r^2 + a^2 \; \cos ^2 \theta$ and $ \; \Delta = r^2-2Mr + a^2$.
Kerr's metric  has
two integration constants: the mass $M$, and the angular momentum per unit mass $a$. We will only consider
the case $M>0$, and
we will take $a \geq 0$
without loss of generality, since for  $a<0$ we can always change coordinates $\phi \to -\phi$, under which $a \to -a$. Setting $a=0$ (no rotation)
gives Schwarzschild's solution. The $\Sigma=0$  ring singularity at $r=0, \theta=\pi/2$
 is hidden behind a black hole horizon
{\em as long as $a \leq M$}. The inner and outer horizons are located
at the zeroes of $\Delta$:  $\ri   =  M - \sqrt{M^2-a^2}$
and $\ro   =  M + \sqrt{M^2-a^2}$, which 
are just coordinate
 singularities  in (\ref{kerr}). 
Kerr's space-time can  be extended  through these horizons, and new regions isometric
to $I: r > \ro$, $II: \ri < r < \ro$ and $III: r < \ri$ arise ad infinitum, giving rise
to the Penrose diagram displayed in Figure 1.
 According to the Carter-Robinson theorem and further results by Hawking and Wald,
(see, e.g., \cite{carter} and references therein)  if $({\cal M}, g_{ab})$ is an asymptotically-flat stationary
 vacuum black hole  that is non-singular on and outside
its event horizon, then it must be  a $0 \leq a  < M$  member of the two-parameter Kerr family.
In other words, the metric (\ref{kerr}) with $a<M$ is the  {\em unique} rotating black hole
solution of Eintein's equations. All these features, together with
 its relevance to model astrophysical
processes, signal Kerr's metric as one of the most important among the known exact solutions of Einstein's equations. \\

\begin{figure}[h]
\centerline{\includegraphics[height=9cm]{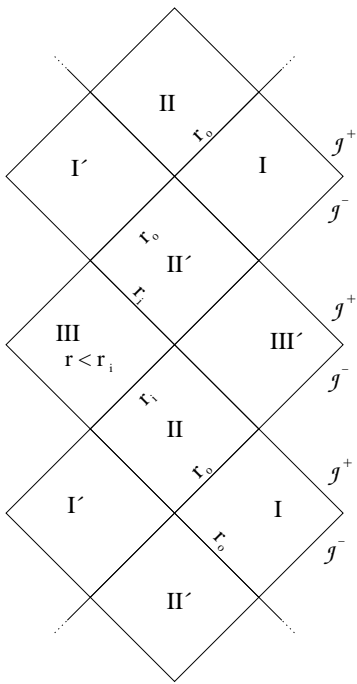}}
\caption{Penrose diagram for the maximal analytic extension of Kerr's space-time. Regions labeled
I and  I' are isometric, and so are II and  II', and III , III'}
\end{figure}

 Regions I and III of a Kerr black hole are stationary, and finding out whether or not they are
 stable under gravitational perturbations is a fundamental issue.
The stability of region I was established after a series of papers following
Teukolsky's discovery of the perturbed  fields  admitting separable evolution equations
\cite{teukolsky}, and ending with Whiting's proof that no solution of  Teukolsky
 equations, supported in region I and properly decaying at infinity and
$\ro$,
 can grow exponentially in time \cite{stable}. According to these results,
 small disturbances
 constrained to region I will oscillate around (\ref{kerr}), in the same way a
 mechanical system oscillates around a local minimum configuration of its potential energy. Region III has a number of undesirable properties: it admits  closed time like curves, has the ring singularity, and  lies beyond the Cauchy horizon, so that predictability
  is lost for earlier data.  It has been argued \cite{pi} that a ``mass inflation'' mechanism would
  prevent the formation of the Cauchy horizon, that would then  be replaced by a null singularity acting as a boundary of
the   spacetime and cutting off this region.
A fundamental issue to address is then the analysis of the linear stability of this region under gravitational perturbations.
In  this paper we show that region III is {\em unstable}
 by proving that there are  solutions of Teukolsky  equations, restricted to region III and satisfying appropriate
boundary conditions, that grow exponentially in time.
This instability
of Kerr space-time remained uncovered since Kerr's metric discovery
 forty five years ago and   makes  the analytic extension of Kerr's
 solution less relevant.
One
 is immediately led to ask where is region III driven to by a gravitational perturbation that excites the unstable modes.
Unfortunately, answering this question goes beyond the linear
perturbation  analysis in this paper.
\\

We also prove in this paper
that (\ref{kerr}) with $a>M$
is unstable.
For this ``super extremal" case,  the ring singularity at $r=0, \theta=\pi/2$
 is {\em naked}, i.e.,
 not protected by a
 black hole horizon, and  can therefore causally connect with the rest of space-time.
 The super extremal case was considered recently in \cite{doglepu}, where numerical evidence for the existence of unstable modes was presented. Because of the nature of the numerical procedures used, the evidence was restricted to a range of values of $a/M$. Clearly, a more general result, establishing in a  rigorous way the unstable nature of the Kerr space time in the case $a>M$ is of interest, as it  suggests that such space times cannot be the end point of the collapse of a self gravitating system, a result of relevance in relation to the Cosmic Censorship Conjecture.
For other nakedly singular space times such as the  negative mass Schwarzschild solution and super extreme Reissner-Nordstr\"om spacetimes,
the linear instability
was established after \cite{dottigleiser,ghi} and \cite{doglepu,cardoso} respectively, where explicit  expressions for unstable modes
can be found.
It should be stressed, however, that
$a>M$ Kerr space-times  are a subset of a larger family of axially symmetric solutions with a
naked singularity, not singled out by the uniqueness theorems that make the black hole
case so relevant.
\\

 This paper is organized as follows: Section \ref{te}  reviews Teukolsky radial and angular equations, focusing
 on axisymmetric unstable modes. Although all we need from the angular  equation is the behaviour of eigenvalues
   for small and large purely imaginary frequency, available  in the recent  literature \cite{bertilong}, a technique to obtain approximations
    for   eigenvalues and eigenfunctions is developed for completeness, and tested against the results in \cite{bertilong}.
    The reformulation of the radial Teukolsky equation that
 follows is crucial in this work. The radial equation is cast in Schr\"odinger like form, and its potential  analyzed.
 This formulation is used in Section \ref{proof} to prove that there exist infinitely many unstable modes for the super-extremal Kerr
 spacetime. Finally, in Section \ref{proof2}, we adapt this proof to show that region III of  a Kerr
 black hole is also unstable.

\section{Teukolsky equations \label{te}}

The systematic analysis of the linearized perturbations of Kerr space-time was greatly facilitated by Teukolsky's discovery \cite{teukolsky} that, in the Newman-Penrose formalism, one can write second order partial differential evolution equations for two linearized null tetrad components $\Phi_s$ of the Weyl tensor, with  $s=\pm 2$ the spin weight of these components. Furthermore these equations
are separable.
Introducing
\begin{equation}
\Phi_s=
R_{\w,m,s}(r) S^m_{\w,s}(\theta) \exp(im\phi) \exp(-i\omega t),
\end{equation}
 the linearized equations reduce to  a coupled system for $S$ and
$R$,
\begin{eqnarray} \label{ta}
{1\over \sin\theta} {d \over d\theta}\left(\sin\theta {d S\over
d\theta}\right)+\left(a^2\omega^2\cos^2\theta-2 a \omega s \cos\theta -{(m + s \cos\theta)^2\over
\sin^2\theta}  +E -s^2\right)S &=& 0 \\
\label{tr}
\Delta {d^2 R \over dr^2} +(s+1) {d\Delta\over dr} \;{dR\over dr} +\left\{
{K^2-2is(r-M)K\over \Delta}+4ir\omega s -[E-2am\omega+a^2
\omega^2-s(s+1)]\right\}R &=&0 ,
\end{eqnarray}
where $\Delta=r^2 -2 Mr +a^2$, and $K=(r^2+a^2)\omega -a
m$, with $M$ the mass, and $a$ the rotation parameter, related to the angular momentum $J$ by $J=aM$.
  Teukolsky's equations also apply to scalar ($s=0$) and electromagnetic
($s=\pm1$) perturbations on the Kerr background.
Moreover, any solution $\psi$ of the $s=-2$
Teukolsky equations is a ``Debye potential'' from which a metric
perturbation $h_{ab}$ that solves the linearized Einstein equations
around the Kerr background can be constructed,  $h_{ab}$ being obtained
by applying a second order linear differential operator to $\psi$
(see \cite{chirinovski}). Teukolsky's equations have been applied in a wide range of problems related to the behavior of Kerr black holes under gravitational and other perturbations, including black hole collisions in the so called ``close limit" \cite{cl}.
 A very important question addressed to from the early applications is that of the stability of region I of a  Kerr black hole under
  gravitational perturbations, which was established through a series of works starting from \cite{teukolsky} and ending
  with \cite{stable}.
In what follows we will  prove that region III  of an
$a<M$ Kerr spacetime,  as well as
 the nakedly singular $a>M$ Kerr  spacetimes,
 are unstable. This is done by  showing
 that for these cases there exist solutions of the Teukolsky equations with purely imaginary  frequency $\omega$ and
 satisfying appropriate boundary conditions.
 Our proof is based on an analysis of the behavior of the lowest  eigenvalues $E$ of (\ref{ta}) and (\ref{tr}) for
large  purely imaginary  frequency $\omega$.
Note that the coupling in the system (\ref{ta})-(\ref{tr}) is rather
subtle, and comes from this common eigenvalue $E$.
Suppose $s$ and $m$ are fixed, then given $\w$, $E$ has to be chosen so that $S$ is regular on the sphere (such a solution
is called a {\em spin weighted spheroidal harmonic, SWSH}),
 and such that  $R$ satisfies the required  boundary conditions. The first condition gives a countable
 set of possible values $E_{\ell}(a \w)$, whereas the second one fixes the possible $\w$ values.
 Each one of these conditions can be cast in the form of  a continued fraction equation \cite{leaver} that arises when
 solving the three term recursion relation on the coefficients of series solution for
  $S$ and $R$ in (\ref{ta}) and (\ref{tr}).

 \subsection{Spin weighted spheroidal harmonics}
  For complex $\w$ near $\w=0$, a Taylor expansion for $E_{\ell}(a \w)$
 up to order $(a \w)^6$ is available in the literature (\cite{seidel}, \cite{bertilong} and references therein). Asymptotic expansions
 for $\w \to \infty$ and for $\w \to +i \infty$ can be found in \cite{lnp,brw,bertishort,bertilong}.
 Particularly useful to our purposes is the fact that
  in the axial case $m=0$, and for gravitational perturbations $s= \pm 2$,
  the $E$ eigenvalue for  large purely imaginary $\w$
  behaves as \cite{bertilong}
 \begin{equation} \label{arrivaberti}
 E_{\ell}(a\w)|_{a\w=ik} = (2 \ell -3) k + {\cal O} (k^0), \;\;\; \text{ as }  k \to \infty
 \end{equation}
In the above formula it was assumed that
 that  the $\ell$ labeling is chosen such that the $(\ell,m,s)$ SWSH
reduces to the corresponding spin weighted {\em spherical} harmonic in the $a \w=0$ limit \cite{bertilong},
which implies
\begin{equation} \label{abajoberti}
E_{\ell}(a\w)|_{a\w=0} = \ell(\ell+1) ,
\end{equation}
the relevant modes being $\ell=2,3,4,..$.

For the purposes of this paper we set $m=0, a\w=ik,s=-2$, and, after changing variables to $x=\cos \theta$ in (\ref{ta}), we find that the angular Teukolsky equation is
\begin{equation} \label{ta3}
 \frac{d}{dx}\left[(1-x^2) \frac{d S}{
d x}\right] - \left(  k^2 x^2 + \frac{4}{1-x^2} \right)S + i \;  4   k  x S = -E S .
 \end{equation}
 If $k=0$, (\ref{ta3}) reduces to an associated Legendre equation, whose regular solutions
are the associated Legendre polynomials $P_{\ell}^{(2)}(x), \ell =2,3,...$, for which
 (\ref{abajoberti}) holds.
For real nonzero $k$,  (\ref{ta3}) is a complex equation, but still the allowed $E$ values
are real (see e.g.,equation (2.3) in \cite{bertilong}). The symmetries of  (\ref{ta3}) for real $E$
imply that if $S(x)$ is a  solution  then so are  $\overline{S(-x)}$
and $S_{\pm}(x) := S(x) \pm \overline{S(-x)}$ (an overline denotes complex conjugation), so we can restrict our search to solutions with
real and imaginary parts of
opposite parity and, multiplying by $i$ if necessary, we can always
 assume that $S(x)$ has even real and odd imaginary parts \footnote{Note that, with this overall
  phase choice, in the $k \to 0$ limit $S$ will reduce to a real multiple of $P_{\ell}^{(2)}$ ($i P_{\ell}^{(2)}$)
  if $\ell$ is even (odd). Note also that our phase choice is different from the choices  in \cite{doglepu} and \cite{bertilong}}:
\begin{equation} \label{parity}
  S(x)= \overline{S(-x)}
\end{equation}
Equation (\ref{ta3}) has regular singular points at $x=\pm1$. There is a  regular Frobenius solution around
$x=-1$ of the form ${\cal F}_{(-1)}(x)=\sum_{n \geq 1} c_n(E,k) \; (x+1)^n$, and a  linearly independent  solution diverging as $\ln (x+1)$.
 The  series in the regular solution converges for $|x+1|<2$ \cite{ms}, and is finite (in fact, vanishes) at $x=1$
if $E=E_{\ell}(a\w)|_{a\w=ik}$ for some $\ell$. Note that in this  case ${\cal F}_{(-1)}(x)$ has to agree up to a multiplicative constant
with the (regular) Frobenius series solution for (\ref{ta}) around $x=1$.
  A SWSH is a solution of (\ref{ta3}) which is
  regular  for $x \in [-1,1]$. Regularity in $[-1,1]$ will follow from regularity at, say,
  $x=-1$ if we impose the  parity condition (\ref{parity}).
Note, however, that  since $S$ satisfies a second order ODE, the condition (\ref{parity}) is equivalent
  to   $\Im (S(0)) = \Re (S'(0)) = 0$.
In particular, if
 $S=(\overline{{\cal F}_{(-1)}(0))} \; {\cal F}_{(-1)}(x)=\sum_{n,m \geq 1} c_n(E,k) \overline{c_m(E,k)} (x+1)^n$, then
$\Im (S(0))=0$, the  solutions $E(k)$ of the equation
\begin{equation} \label{root}
0 = \Re (S'(0)) = (1/2) \sum_{n,m \geq 1} n  (c_n(E,k) \overline{c_m(E,k)} + \overline{c_n(E,k)} c_m(E,k)) =: {\cal D}(E,k),
\end{equation}
will  equal $E_{\ell}(a \w)|_{a \w=ik}$ for some $\ell$, and
$S$ will be  regular for $x \in [-1,1]$ for these $E$ values.\\
As an application of this approach, we replaced ${\cal F}_{(-1)}(x)$ in (\ref{root}) with the regular Frobenius series
at $x=-1$ up to order $(x+1)^{40}$, and numerically solved for the lowest $E$ (i.e., $\ell = 2$)
root in (\ref{root}) for different
values of $k$. The results, contrasted with the low $\w$ expansion
(see  \cite{seidel} \cite{bertilong}
and references therein) and the asymptotic form (\ref{arrivaberti})
  are  shown in Figure 2.\\
\begin{figure}
\centerline{\includegraphics[height=7cm]{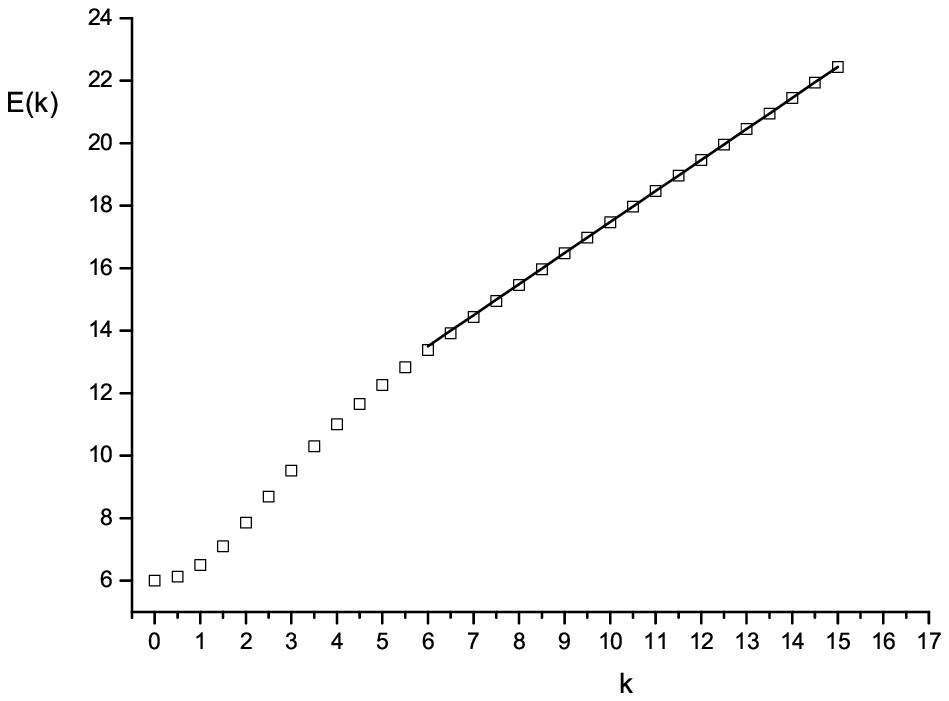}
\includegraphics[height=7cm]{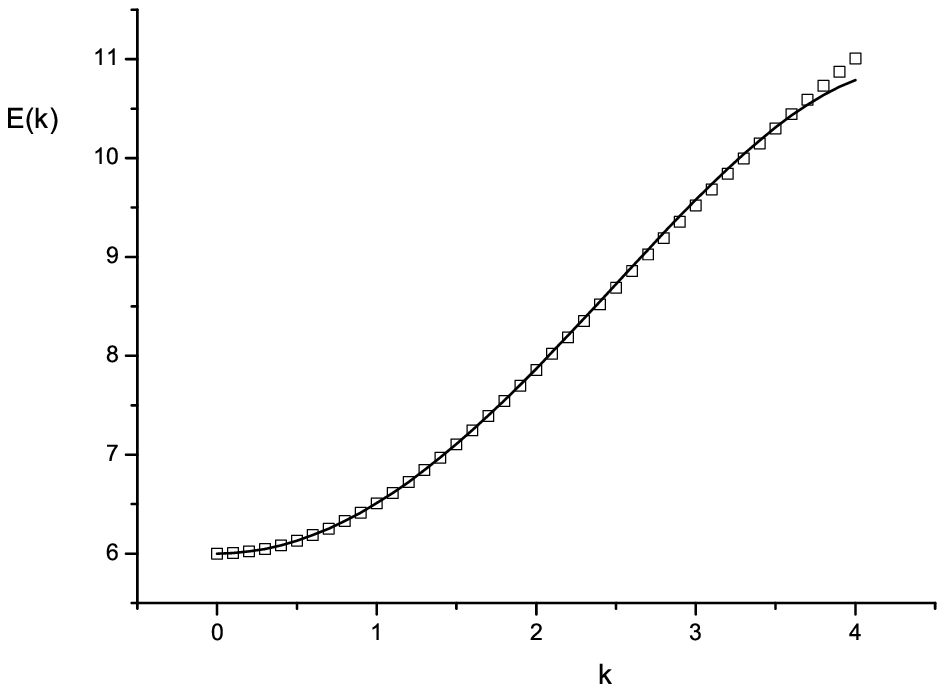}}
\caption{The left panel shows the lowest (corresponding to $\ell=2$) $E$ roots obtained by solving eq. (\ref{root}) for thirty
different values of $k \in [0,15]$, and a least square linear fit for the ten points in the interval $10\leq k \leq 15$,
 which gave $E=0.993k+$constant, in excellent agreement with the expected asymptotic expansion
 Eq. (\ref{arrivaberti}). The right panel shows forty values obtained in
this way for $0<k<4$ (dots) and the low frequency  approximation in \cite{seidel} \cite{bertilong}
 up to order $k^4$ (solid line).}
\end{figure}

\subsection{Teukolsky radial equation, $a>M$ case}

Equation (\ref{tr}) is of the form  $\Delta \ddot R + Q \dot R + Z R = 0$, dots denoting
derivatives with respect to $r$, and can be
put  in a suitable form for our analysis by introducing an integrating factor $L$,
 $\psi:=R/L$ and changing the radial variable to $r^*$, where $\frac{d r^*}{dr} := \frac{1}{f}$
with  $f$  an unspecified positive definite function of $r$. The result is
\begin{equation} \label{r1}
\frac{\Delta L}{f^2} \psi'' + \left( \frac{ 2 L' \Delta}{f^2} + \frac{Q L}{f} - \frac{\Delta L f'}{f^3} \right) \psi' +
\left(\frac{\Delta L'' }{f^2} + \frac{Q L'}{f} - \frac{\Delta L' f'}{f^3} +Z L \right) \psi = 0.
\end{equation}
where $'$ denotes derivatives with respect to $r^*$. The choice
$f=\sqrt{\Delta}$ (recall $\Delta$ is strictly positive in the super extreme case $a>M$) gives an adimensional $r^*$,
\begin{equation} \label{rs}
r^* = \ln \left( \frac{r-M+\sqrt{r^2-2Mr+a^2}}{M} \right)  \simeq \begin{cases} \ln \left(\frac{2 r}{M} \right) & r \to \infty \\
\ln \left( \frac{a^2-M^2}{2 M |r| } \right) & r \to - \infty \end{cases}
\end{equation}
that grows monotonically with $r$, and  can  easily be inverted in terms of elementary functions,
\begin{equation} \label{irs}
r = \frac{M \exp(r^*)}{2} + M + \frac{M^2-a^2}{2M \exp(r^*)}.
\end{equation}

The next step is to  choose $L$ such that the $\psi'$ coefficient of (\ref{r1}) vanishes. This gives
$L = \Delta ^ {- \frac{(2s+1)}{4}}$ with  (\ref{tr}) reducing to
 a Schr\"odinger like equation
\begin{equation} \label{scho}
{\cal H} \psi := - \psi'' + V \psi = -E \psi.
\end{equation}
The potential
\begin{equation}
V :=  - \left( \frac{\Delta \ddot L}{L} + \frac{Q \dot L}{L} + Z + E \right)
\end{equation}
is independent of $E$ (recall that $Z$ is the coefficient of $R$ in (\ref{tr})),
 this being the reason behind the above choice of $r^*$.
For $s=-2$, $m=0$, and $ \omega=ik/a$, the
 case we are interested in, the potential is real and reads
\begin{multline} \label{v}
V = \left[\frac{r^4+ a^2 r^2 +2 a^2 M r}{a^2(r^2-2Mr+a^2)} \right]  \; k^2 +
 4 \left[ \frac{-r^3 + 3 M r^2-  a^2r-a^2 M}{a(r^2-2Mr+a^2)} \right]  \; k  + \left[ \frac{r^2-2Mr+15 M^2-14a^2}{4(r^2-2Mr+a^2)}
 \right] =: k^2 V_2 + k V_1 + V_o
\end{multline}
Since $a>M$ in the super-extreme case,
 $V$ is smooth everywhere. Also, $V$ is bounded from below for every $k \geq 0$, its minimum
 being not continuous, as a function of $k$, at $k=0$:
 \begin{eqnarray} \label{vm}
 \min \{ V(r,k=0), r \in {\mathbb R} \} & = & -7/2 \\ \nonumber
 \lim_{k \to 0^+}  \min \{ V(r,k), r \in {\mathbb R} \} & = & -15/4
 \end{eqnarray}

\section{\label{proof} Unstable modes of the Kerr naked singularity}

 ${\cal H}$ in (\ref{scho}) is self-adjoint in the Hilbert space of square integrable functions of $r^*$ with
hermitian product  $\langle \alpha | \beta \rangle := \int \overline{\alpha} \; \beta \; dr^*$.
 Given that  $V$ is smooth,  bounded from below, and
 \begin{equation} \label{a1}
 V \sim  \left( \frac{Mk e^{r^*}}{2a} \right) ^2 ,  |r^*| \to \infty,
 \end{equation}
the  spectrum of the self-adjoint
  operator ${\cal H}$  is fully discrete and has a lower bound. A careful analysis
   of (\ref{rs}), (\ref{v}) and (\ref{a1}), or, more directly, the asymptotic form of (\ref{tr}),
   indicates that the square integrable eigenfunctions of ${\cal H}$ behave
as
\begin{equation} \label{ae}
\psi \sim \begin{cases} e^{- \frac{rk}{a}} \left( \frac{M}{r} \right) ^ {\frac{2kM}{a}-3} \left( 1 + {\cal O} (M/r) \right)  & , r \to \infty \\
                    e^{ \frac{rk}{a}} \left( \frac{M}{r} \right) ^ {1-\frac{2kM}{a}} \left( 1 + {\cal O} (M/r) \right) & , r \to -\infty  \end{cases}
\end{equation}
with sub-leading terms that depend on the eigenvalue.\\
To obtain some useful information about the fundamental energy of the radial Hamiltonian (\ref{scho}), we need to analyze
its potential (\ref{v}). The numerator of $V_2$,
$r (r^3+ a^2 r  + 2a^2M )$, is negative in some interval to the left of zero, and, since
 $V_1(0) = -4a^2M <0$, by continuity $V_1$   must also be negative in some neighborhood  of $r=0$.
 Thus, there is  an interval $s_1(M)< r < s_2(M)<0$ where both
 $V_1$ and $V_2$ are negative.
Let $\psi$ be a
smooth normalized ( $\langle\psi|\psi\rangle=1$) test function
supported on
$r \in (s_1(M),s_2(M))$. Note  that
  $\psi$ has  compact support away from $r=0$, and that
\begin{equation} \label{test1}
\langle \psi | {\cal H} | \psi \rangle =   \langle \psi | - ( \partial / \partial r^* ) ^2 |
\psi \rangle +
\sum_{j=0}^{2} k^j \langle \psi |V_j |\psi \rangle
\end{equation}
with
\begin{equation} \label{test2}
\langle \psi |V_j |\psi \rangle =
 \int_{s_1(M)}^{s_2(M)} | \psi |^2 \; V_j \; \frac{dr}{\sqrt{\Delta}} < 0
\text{ for } j=1,2.
\end{equation}
Since $\langle \psi |V_1 |\psi \rangle$  is negative, there is a $k_c$ such that,
for $k>k_c$,
$ \langle \psi | - ( \partial / \partial r^* ) ^2 |
\psi \rangle +
\langle \psi |V_0 |\psi \rangle + k \langle \psi |V_1 |\psi \rangle $ is  negative
and so $\langle \psi | {\cal H} | \psi \rangle  <  k^2  \langle \psi |V_2 |\psi \rangle$.
This implies that if $-\epsilon_o(k)$
is the lowest  eigenvalue of ${\cal H}$ then
\begin{equation} \label{br}
 -\epsilon_o(k) \leq \;  \langle \psi | {\cal H} | \psi \rangle  <  k^2  \langle \psi |V_2 |\psi \rangle  < 0 \, , \; \; \text{ if } k > k_c.
\end{equation}
From (\ref{vm}) and the above equation
\begin{equation} \label{radiale}
\epsilon_o(k=0^+)   <  \frac{15}{4}, \hspace{1cm}
 \epsilon_o(k) > | \langle \psi |V_2 |\psi \rangle | k^2 , \; \; k > k_c.
\end{equation}
From (\ref{arrivaberti}), (\ref{abajoberti}) and (\ref{radiale}) it is clear that, for any
$\ell=2,3,...$, the curves $\epsilon_o(k)$ and  $E_{\ell}(a\w)|_{a\w=ik}$ intersect at some
$k_{\ell} > 0$. We have numerically checked this behavior  and spotted  the predicted intersection point $k_{(\ell=2)}$.
 The results, shown in Figure 3, are
in perfect agreement with those in \cite{doglepu}.
\begin{figure}
\centerline{\includegraphics[height=7cm]{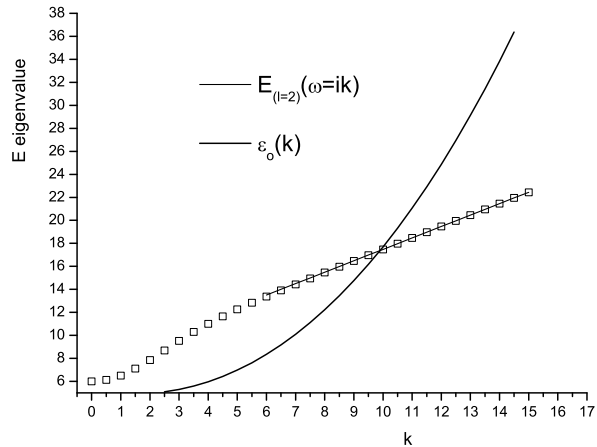}}
\caption{Intersection of the numerically generated curves $E_{(\ell=2)}(a\w)|_{a\w=ik}$ and $\epsilon_o(k)$ for $a/M=1.4$.
Note the agreement with the results in \cite{doglepu}, according to which  for $a=1.4$ the intersection should occur at $k \simeq
7.07 \times 1.4  = 9,898$.}
\end{figure}

 This implies that there is a common eigenvalue $E =  E_{\ell}(a\w)|_{a\w=ik_{\ell}}
=  \epsilon_o(k_{\ell})$ for the radial and angular Teukolsky equations and that
 $\{ S(\theta)=S_{(\ell,m=0,k_o)}(\theta), R(r) =
\Delta ^ {\frac{3}{4}} \psi_o^{(k_{\ell})}(r) \}, \ell=2,3,..$ are (infinitely many!)
solutions of the Teukolsky equations, all of which correspond to
 unstable gravitational  perturbations
$ \Phi_{(s=-2)}(t,r,\theta)=S_{(\ell,m=0,k_o)}(\theta)  \Delta ^ {\frac{3}{4}} \psi_o^{(k_{\ell})}(r) \exp(k_{\ell} t/a)$ with
$\psi_o^{(k_{\ell})}(r)$ behaving as in (\ref{ae}).
The radial exponential  decay rate guarantees that any quantity of interest
vanishes asymptotically for large $r$. \\
Regarding the behavior of the perturbation
at the ring singularity we point out that: (i) Contrary to what happens for
other naked singularities (e.g., negative mass Schwarzschild solution, see \cite{ghi,dottigleiser}),
$r=0$ is {\em not} a singular point of the radial equation, the potential (\ref{v})
being smooth everywhere. In fact,
$r=0$ is neither a boundary point,
nor a singular point of the second order radial equation (\ref{scho}),
then imposing  conditions at this point,
besides requiring that the perturbations vanishes at
 $r^* = \pm \infty$, would lead to an unnaturally
 over-determined problem. (ii) It is natural to ask, however,
 what is the effect of the perturbation on the singularity.
 In particular, one wants to make sure that
 first order  corrections to the Riemann tensor invariants do not diverge faster than the
  zeroth-order term as we approach the singularity, otherwise,
  the use of linearized equations is not justified, since we cannot keep the perturbation small in the whole spacetime. As shown in \cite{doglepu}, no algebraic
  invariant gets
  corrections under our unstable modes.
 A deeper treatment of the perturbed geometry could possibly be afforded in the approach where
 the  $s=-2$ solutions of the  Teukolsky
equation  are used as ``potentials" to construct  {\em metric} perturbations \cite{chirinovski}, but this is a rather technical analysis \cite{futuro} that
  certainly exceeds the scope of this report.

 Note that the unstable perturbations are
 not restricted to the fundamental radial mode, since for
  large $k$, the potential $V$ in (\ref{scho}) has a deep minimum
   that can be approximated by a harmonic oscillator potential of depth of order $k^2$, and
   strength also of order $k^2$, so that the level spacing near the ground state of  ${\cal H}$ is
   of order $k$. Thus, the negative   $-\epsilon_n(k)$ of the lowest radial eigenvalues
 grow also quadratically in $k$  for large $k$, and  intersect the angular eigenvalue curves for
 sufficiently large values of $k$ \cite{futuro}.\\

\section{\label{proof2} Unstable modes for region III of a Kerr black hole.}
The calculations  above can be easily adapted to deal with perturbations in the {\em interior} region $r< \ri:= M - \sqrt{M^2-a^2}$
of an $a \leq M$ Kerr {\em black hole}.  The extreme $a=M$  and
 sub-extreme $a<M$ cases require separate treatments. \\

 \noindent
\subsection {Extreme case (a=M) inner region}
 The solution
 of $dr^*/dr = 1/\sqrt{\Delta}$ in the interior region  $r < \ri = \ro = M$ is
 \begin{equation} \label{rse}
r^* =
-\ln \left( \frac{M-r}{M}  \right), r < \ri,
\end{equation}
with inverse
\begin{equation} \label{irse}
r = M ( 1- e^{-r^*}) , \;\;  -\infty < r^* < \infty
\end{equation}
Using  the
  integration factor $\Delta^{3/4}$ as before, we are led back to (\ref{scho}) and (\ref{v}), with
 $r$ given in (\ref{irse}).
  Note that
 \begin{equation} \label{av2}
V \sim \begin{cases}  4k^2 \exp(2r^*)  & , r^* \to \infty \\
                k^2 \exp(-2r^*)     & , r^* \to -\infty , \end{cases}
\end{equation}
then the   spectrum of the self-adjoint
  operator ${\cal H}$  is again fully discrete and has a lower bound. The eigenfunctions behave as
 \begin{equation} \label{aee}
\psi \sim \begin{cases} \left(\frac{M}{M-r} \right)^ {2k} \; \exp \left[-2k \left(\frac{M}{M-r} \right) \right]
 \left( 1 + {\cal O} (\frac{M-r}{r}) \right)  & , r \to M^- \\
                    \frac{M}{r}^{1-2k} \; e^{ \frac{rk}{M}}  \left( 1 + {\cal O} (M/r) \right) & , r \to -\infty  \end{cases}
\end{equation}
The argument of instability in the super extreme case goes through in this case without modifications, since the test function in (\ref{test1}) is supported in the $r<0$ region,
and thus can be used again in this case to obtain the bound (\ref{br}).
The radial decay (\ref{aee}) guarantees that corrections to relevant quantities vanish in these limits.
The considerations
about  the perturbation behavior near the ring singularity for the super extremal case apply
again  to this case, and also to the sub-extreme case below. \\

\subsection{Sub-extreme case ($a < M$) inner region}
The sub-extreme case introduces some subtleties: requiring $dr^*/dr = 1/\sqrt{\Delta}$ for  $r< \ri$ gives
\begin{equation} \label{urs}
r^* = \ln \left( \frac{M-r-\sqrt{r^2-2Mr+a^2}}{M} \right)
 = \ln \left( \frac{\ri + \ro-2r-2 \sqrt{(\ro-r)(\ri-r)}}{\ri+\ro} \right), \;\; r< \ri ,
 \end{equation}
and the radial equation reduces to (\ref{scho}), (\ref{v}) with
$r$ the inverse of (\ref{urs}). \\
Since
\begin{equation}
-\infty < r^* < \ri^* :=  \ln \left( \frac{\ro - \ri}{\ro+\ri} \right) ,
\end{equation}
 (\ref{scho}) is, in this case,  a
Schr\"odinger like equation {\em on a  half axis}, with
  potential
     diverging  as $$V \sim [k (M^2-a^2)/(2aM)]^2~\exp(-2 r^*)$$ for  $r^* \to -\infty$
    and, for  $r^* \to \ri^*{}^-$, as
 \begin{equation}  \label{sqp}
V \sim \left[ -\frac{1}{4} + \frac{4M^2 \left( M-\sqrt{M^2-a^2} \right) ^2}{M^2-a^2} \left(
\frac{k}{a} - \frac{\sqrt{M^2-a^2}}{M \left( M - \sqrt{M^2-a^2} \right)} \right)^2 \right] \frac{1}{(\ri^*-r^*)^2} + \cdots =: \frac{\nu(k)^2-\frac{1}{4}}{(\ri^*-r^*)^2}+ \cdots ,
\end{equation}
 $\nu(k) > 0$.
 The subtleties mentioned above are the half infinite domain for $r^*$ and the  behaviour (\ref{sqp}) of the potential
 at the $\ri^*$ boundary. These singular potentials are  discussed in detail in \cite{meetz}.
 The eigenfunctions of ${\cal H}$   which are square integrable for
  $r^*$ near   $ - \infty$ behave as in (\ref{ae}) in this limit and as
\begin{equation}
\psi \sim   a \left[(\ri^*-r^*)^{\frac{1}{2}+ \nu} + ... \right] + b  \left[ (\ri^*-r^*)^{\frac{1}{2}- \nu} + ... \right] ,  \label{zero}
\end{equation}
near the horizon,
 the $E$ eigenvalue appearing only in sub-leading terms.
  Thus, if  $\nu>1$,
for  generic $E$ these are   not square integrable near zero, unless $b=0$,
and this is precisely the condition that
selects a discrete set of possible $E$ values as the spectrum of ${\cal H}$, and that defines de space of functions
where ${\cal H}$ is self-adjoint.
On the other hand, if $\nu<1$, {\em for any $E$} the eigenfunction behaving properly at minus infinity will be  square integrable
for $r^* \in (-\infty,0)$, and a choice of boundary condition needs to be imposed to define
a set of allowed perturbations  $D_{phys}$,
in order that  ${\cal H}$ be a self-adjoint operator on $D_{phys}$ (and thus have
 a complete set of eigenfunctions) \cite{reed}.
This situation, where a choice of self-adjoint extension needs to be done,
 is also found in the study of perturbations around the negative mass Schwarzschild
space-time \cite{ghi}, and perturbations in the inner region of a Reissner-N\"ordstrom
space-time \cite{futuro}. Roughly,  $D_{phys}$ is constructed by adding to the
 set of functions of compact support, those
decaying as (\ref{zero}) with a {\em fixed} quotient $q:=b/a \in {\mathbb R}$, possibly infinite
\cite{reed}.
 However,
regardless of our choice for $q$, functions of compact support away from this singular
point will
belong to  $D_{phys}$, in particular, the test function used in
(\ref{test1})-(\ref{test2}). This implies that
the bound  (\ref{br}) will hold for any choice of boundary condition ($q$ value)!
In other words,  region III of a sub-extreme Kerr black hole is unstable regardless
the boundary condition imposed to perturbations at the inner horizon.
A natural choice would be that giving the fastest decay, i.e., $q=0$. In this case
 the gravitational perturbations behave as
$\Phi \sim \Delta^{3/4} (\ri^*-r^*)^{\frac{1}{2}+ \nu} \sim (\ri^*-r^*)^{2 + \nu} \sim (\ri-r)^{1 + \nu/2}$.
One should keep in mind that  $\nu = \nu(k_{\ell})$ is positive and grows for large $\ell$.\\

The instability found in region III of Kerr space-time, and the fact that
the  Reissner-N\"orsdtrom charged black hole  also has a two horizon
structure with an inner static region $r<\ri$, triggers the question of
whether of not this inner static region is stable.
Preliminary work indicates
that this region is unstable \cite{futuro}.\\

As a final comment, we note that our results do not contradict the established
stability of the {\em outer} stationary, region I of Kerr black holes.
For $a \leq M$ and $r>\ro$ we can define $r^*$ as in (\ref{rs}), and get again
the Schr\"odinger form (\ref{scho}), but now the domain
is restricted to $\ro < r$, where
\begin{equation} \label{mierda}
V > -\frac{15}{4} \; \text{ if } a \leq M \; \text{ and } \;  0 < k.
\end{equation}
This implies that $\epsilon_o(k) < 15/4$ for all $k>0$ and thus cannot intersect
$E_{\ell}(a\w)|_{a\w=ik}$ for {\em any}  $\ell$. To get the bound (\ref{mierda}) is
easy in the extreme case, for which (\ref{v}) reduces to (replace $r=M r'$, then drop the primes)
\begin{equation} \label{ve}
V_{ext} = \left[ \frac{r(r+1)(r^2-r+2)}{(r-1)^2} \right] k^2 - 4 \left[ \frac{r^2-2r-1}{r-1} \right] k  +\frac{1}{4} .
\end{equation}
We search  for critical points at fixed $k$ by solving $\partial V / \partial r =0$. This gives
\begin{equation} \label{kr}
k = \frac{2(r-1)(r^2-2r+3)}{(r^2+1)(r^2-2r-1)},
\end{equation}
which has  a unique $r>1=\ri=\ro$ solution for every positive $k$, with $r$ moving from infinity down to
$1 + \sqrt{2}$ as $k$ goes from zero to infinity.
Since $\partial^2 V / \partial r^2 >0$ at these points, these are (global)
minima of $V$ {\em for fixed $k$}. Moving along this curve from $r= 1 +\sqrt{2}$ to infinity,
$V$ decreases monotonically to $-15/4$. In the sub-extreme case $a<M$ the equation $\partial V / \partial r =0$
is quadratic in $k$, and yields two curves like that in (\ref{kr}). However, one of these curves gives $k<0$ for
all $r>r_{ext}$, and should be discarded. The other one corresponds to minima of $V$ for fixed positive $k$, with
$V$ again decreasing monotonically to $-15/4$ along it.

\section*{Acknowledgments}

We thank Vitor Cardoso and Marc Casals for clarification of a number of issues
related to prolate SWSHs.
This work was supported in part by grants from CONICET (Argentina)
and Universidad Nacional de C\'ordoba.  RJG and GD are supported by
CONICET. IFRS is a fellow of the CICBA.


\begin{thebibliography}{99}
\bibitem{kerr} Kerr, R. P., {\em Phys. Rev. Lett} {\bf 11}, 237 (1963).
\bibitem{carter} B. Carter, {\em Has the black hole equilibrium problem been solved?
}, Invited talk at 8th Marcel Grossmann Meeting, Jerusalem, Israel, 22-27 Jun 1997.
e-Print: gr-qc/9712038.
\bibitem{teukolsky} S.~A.~Teukolsky,
  Phys.\ Rev.\ Lett.\  {\bf 29} (1972) 1114;
 Astroph. J. {\bf 185}, 635 (1973).
 \bibitem{stable}
   B.~F.~Whiting,
  J.\ Math.\ Phys.\  {\bf 30}, 1301 (1989).
  \bibitem{pi}
  E.~Poisson and W.~Israel,
  Phys.\ Rev.\  D {\bf 41}, 1796 (1990).

\bibitem{doglepu}
 G.Dotti, R. J. Gleiser, J. Pullin
Phys.Lett. {\bf B644} (2007) 289-293 [arXiv:gr-qc/0607052]
\bibitem{dottigleiser}
R.~J.~Gleiser and G.~Dotti,
  Class.Quant.Grav.{\bf 23}, 5063 (2006) [arXiv:gr-qc/0604021]
\bibitem{ghi} Gibbons G W, Hartnoll D and  Ishibashi A 2005 {\em Prog.Theor.Phys.}
{\bf 113} 963-978, hep-th/0409307.
\bibitem{cardoso}
  V.~Cardoso and M.~Cavaglia,
   Phys.Rev. {\bf D74},   024027 (2006)
  [arXiv:gr-qc/0604101]
\bibitem{leaver}
  E.~W.~Leaver,
  Proc.\ Roy.\ Soc.\ Lond.\  A {\bf 402}, 285 (1985).
\bibitem{seidel} E. Seidel, Class. Quantum Grav. 6 (1989) 1057.
\bibitem{bertilong}
  E.~Berti, V.~Cardoso and M.~Casals,
  Phys.\ Rev.\  D {\bf 73}, 024013 (2006)
  [Erratum-ibid.\  D {\bf 73}, 109902 (2006)]
  [arXiv:gr-qc/0511111].
  \bibitem{chirinovski} L. Kegeles, J. Cohen, Phys. Rev. {\bf D19} ,1641 (1979);
P. Chrzanowski, Phys. Rev. {\bf D11}, 2042 (1975); R. Wald, Phys.
Rev. Lett. {\bf 41}, 203 (1978).
\bibitem{cl} See, e.g., G. Khanna, Phys. Rev. {\bf D66} 064004 (2002)
\bibitem{lnp} R. A. Breuer, Gravitational Perturbation Theory and
Synchrotron Radiation, Lecture Notes in Physics,
Vol. 44 (Springer, Berlin, 1975).
\bibitem{brw} R.A. Breuer, M.P. Ryan, Jr., and S. Waller, Proc. R. Soc. London
A358, 71 (1977).
\bibitem{bertishort}
  E.~Berti, V.~Cardoso and S.~Yoshida,
  Phys.\ Rev.\  D {\bf 69}, 124018 (2004)
  [arXiv:gr-qc/0401052].
\bibitem{ms} J Meixner, FW Sch\"afke, Mathieusche Funktionen und Sph\"aroidfunktionen, 1954 - Springer-Verlag
; Mathieu functions and spheroidal functions and their mathematical foundations: Further studies,
Lectures Notes in Mathematics, vol 837 (1980) Springer-Verlag.
\bibitem{futuro} Dotti et al, work in progress.
\bibitem{meetz} K. Meetz, Il Nuovo Cimento {\bf XXXIV} 690 (1964).
\bibitem{reed}
M. Reed and B. Simon,
{\em Methods of Modern Mathemaical Physics II: Fourier Analysis, Self-Adjointness}, Academic Press, section X.1.

\end{thebibliography}
\end{document}